# Hybrid spectral-spatial domain registration for nanometric tracking in digital in-line holographic microscopy

KAMIL KALINOWSKI*, MIKOŁAJ ROGALSKI, PIOTR ARCAB, EMILIA WDOWIAK, PIOTR ZDAŃKOWSKI, AND MACIEJ TRUSIAK**

*Institute of Micromechanics and Photonics, Warsaw University of Technology, 8 Sw. A. Boboli St., 02-525 Warsaw, Poland*
**kamil.kalinowski2.dokt@pw.edu.pl, **maciej.trusiak@pw.edu.pl*

**Digital in-line holographic microscopy enables label-free tracking and metrology, but achieving nanometric sub-pixel displacement accuracy over a wide capture range remains challenging. Frequency-domain registration based on the discrete Fourier transform (DFT) is globally stable and tolerant to large shifts, yet it suffers from sub-pixel quantization and interpolation artifacts that limit precision near zero displacement. In contrast, spatial-gradient refinement such as Lucas–Kanade (LK) can reach very high sub-pixel accuracy, but it is strongly initialization-limited and prone to divergence outside a narrow convergence basin. Here we propose a Hybrid Spectral–Spatial Domain (HSSD) framework that resolves this trade-off by combining the global robustness of a DFT-based coarse estimator with LK refinement of the residual motion. The DFT stage provides reliable initialization and substantially extends the capture range, while the LK stage suppresses the precision floor by reducing interpolation and quantization errors characteristic of standalone frequency-domain methods. We validate HSSD using numerical simulations and experiments in a transmission in-line holographic imaging system, achieving nanoprecision displacement measurement and stable tracking across a wide range of displacements and defocusing conditions. This hybrid strategy enables reliable nanometric localization in regimes where standalone DFT or LK methods either fail to converge or saturate in accuracy.**

Digital In-line Holographic Microscopy (DIHM) has emerged as a pivotal imaging technique for the high-throughput, three-dimensional analysis of micro- and nano-scale phenomena, from particle tracking and flow analysis to the detection of viruses and biological pathogens [1, 2]. The ability to precisely localize and track sparse objects within the holographic field of view is fundamental to extracting quantitative dynamics from complex biological and physical environments [3, 4, 5]. However, achieving reliable nanometric-scale localization remains a significant challenge, as it requires resolving sub-pixel displacements that often push the limits of classical image registration algorithms.

High-precision measurement of these displacements is a fundamental problem in holographic metrology. Traditionally, frequency-domain approaches based on the Discrete Fourier Transform (DFT), particularly phase-correlation methods, are widely used due to their computational efficiency and robustness to global intensity variations [6,7]. While these methods perform reliably for pixel-level displacements, their accuracy significantly degrades in the sub-pixel regime when dealing with real-world holographic data. In this limit, quantization effects introduce systematic errors that are difficult to eliminate in practice, creating a "precision floor" that hampers nanometric analysis [8].

An alternative class of methods is based on spatial-domain formulations of optical flow, such as the Lucas–Kanade (LK) algorithm [9], which estimates motion by exploiting local intensity gradients. While optical-flow-based approaches are highly effective for resolving infinitesimal sub-pixel displacements, their performance is inherently constrained by a high sensitivity to initial conditions and a relatively narrow capture range. This frequently leads to convergence failures or local minima traps when applied to large or unknown displacements typical of dynamic particle tracking [10].

In this work, we propose the Hybrid Spectral-Spatial Domain (HSSD) approach, which combines the complementary strengths of both domains to overcome these limitations [8].

**Algorithm description.** The proposed estimation strategy is rooted in a complementary analysis of spatial and frequency-domain methods. While frequency-domain approaches based on the DFT are computationally efficient and provide global stability, they suffer a significant loss of sensitivity [11] and "flattening" of the estimator response in the regime of near-zero displacements. Conversely, spatial-domain methods like the LK algorithm exploit local intensity gradients to achieve superior precision for such extremely small sub-pixel shifts. However, as demonstrated in the subsequent numerical analysis, the LK algorithm is inherently limited by a narrow operational range; once the shifts of the recorded holographic patterns exceed approximately 50 pixels, the gradient descent fails to converge. This fundamental trade-off between the global robustness of DFT [12] and the high local sensitivity of LK forms the technical basis for our hybrid framework. As shown in the HSSD workflow in Fig. 1, the process begins with a coarse displacement estimation using a standard DFT-based cross-correlation. The cross-correlation map is computed as $C = \mathcal{F}^{-1}\{F_1 \cdot F_2^*\}$, where $F_1$ and $F_2$ are the DFT spectra of the two input holograms; the integer-pixel displacement estimate $(dx', dy')$ is then obtained as the coordinates of the peak of C. Restricting this estimate to integer values is deliberate: it avoids interpolation artifacts that would otherwise degrade the subsequent refinement. Note that this DFT stage alone yields only pixel-level accuracy; sub-pixel precision is achieved exclusively through the hybrid DFT+LK combination. This stage serves as a robust initialization, effectively bringing shifted holographic patterns into the convergence radius of the LK estimator. Once this global alignment is performed, the LK refinement stage takes over. Instead of registering the entire frame, the algorithm operates on specific feature points detected within the intensity patterns of individual micro-particles. By tracking these local intensity variations - ranging from sharp boundaries of in-focus objects to diffraction patterns in defocused images - the LK stage overcomes the precision bottlenecks and systematic errors typical of frequency-domain methods near the origin. Unlike the DFT stage, which assumes purely translational motion and serves solely as a coarse initializer, the LK refinement naturally accounts for affine deformations of the holographic intensity pattern - including local rotation, scale variation, and shear - through its gradient-based optical flow formulation. To improve robustness against noise, the algorithm tracks multiple feature points distributed across the particle's intensity pattern simultaneously, with each point defined by a local intensity gradient automatically identified by the LK estimator. Feature points are selected using the Shi-Tomasi corner detector with up to 200 candidates, a quality threshold of 0.01, and a minimum inter-point separation of 10 pixels; points for which tracking fails are discarded, with a minimum of 5 valid points required to produce an estimate. The final displacement is determined as the median of all individual point estimates and assembled as $dx = dx' + dx''$ and $dy = dy' + dy''$, where $(dx'', dy'')$ is the sub-pixel residual from LK. Since the DFT stage yields strictly integer-pixel output and its sub-pixel residual constitutes the input signal to the LK stage rather than a source of propagating uncertainty, the measurement uncertainty of the final displacement is characterized by the LK refinement alone and reflected in the combined RMSE reported in Fig. 4.

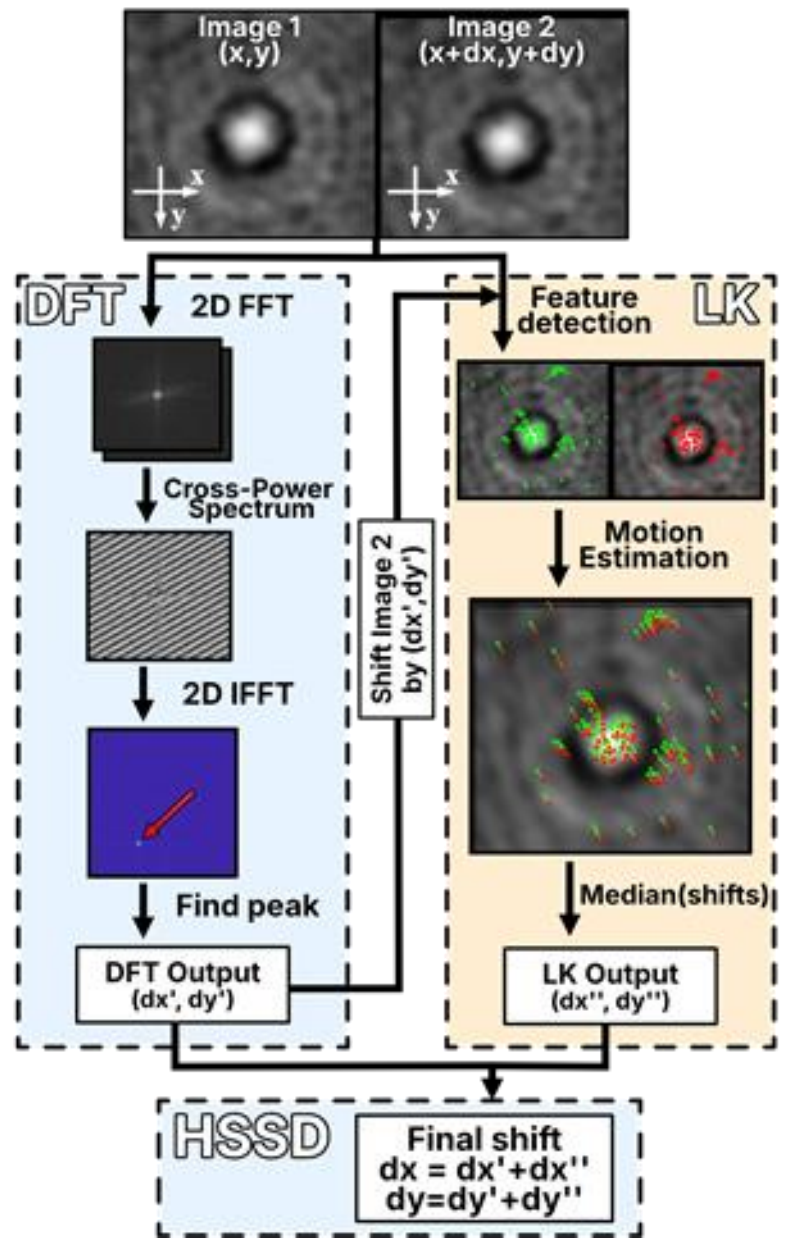

Fig. 1. Schematic of the HSSD algorithm. The DFT stage (blue) computes a coarse, integer-pixel displacement estimate (dx', dy') via cross-correlation in the frequency domain. The LK stage (orange) refines the residual sub-pixel shift (dx'', dy'') using optical flow on feature points detected in the pre-shifted images. The final lateral displacement is given by $dx = dx' + dx''$ and $dy = dy' + dy''$, combining the pixel-level and sub-pixel contributions in the x and y directions, respectively.

**Numerical performance analysis.** To quantitatively evaluate the performance of the displacement estimation methods, we conducted extensive numerical simulations designed to replicate non-ideal imaging conditions. The ground-truth dataset was generated using the experimental hologram of a single microbead particle as a reference. Sub-pixel shifts were introduced by applying affine transformations in the spatial domain, with random displacements drawn from a uniform distribution within the narrow range of ±10 pixels for both the horizontal and vertical axes. A total of 100 such random positions were generated to provide a statistically significant dataset for performance assessment. Six types of image degradation were applied synthetically to each pair of shifted images prior to registration. . Gaussian noise ($\mu = 0$, $\sigma = 1$), median blur (kernel size 5), motion blur (kernel size 5), linear intensity rescaling, Poisson noise ($\lambda$=1) and intensity quantization to 64 gray levels.

The results, as illustrated in Fig. 2, demonstrate that both the HSSD and LK algorithms outperform standard DFT registration in the majority of tested scenarios, showing significantly higher robustness to stochastic noise and blur. However, the analysis also reveals a specific sensitivity of these spatial-domain methods to signal quantization. This performance gap is rooted in the mathematical nature of both the HSSD and LK approaches; while they rely on local spatial intensity variations or gradients, strong quantization transforms smooth transitions into discrete step functions. Such discontinuities lead to unreliable derivative estimations - where gradients either vanish on plateaus or become artificially amplified at edges - thereby degrading sub-pixel precision for both the LK and HSSD methods. In contrast, the standalone DFT-based stage operates in the frequency domain, effectively averaging these quantization effects over the entire image to provide a stable baseline. However, as substantiated by the numerical evidence, the superiority of these spatial refinement methods in terms of precision is strictly conditional. Blur - whether median or motion - acts as a low-pass filter that attenuates high-frequency components and distorts the phase of the DFT cross-correlation peak, directly degrading the accuracy of the frequency-domain stage while leaving spatial gradients largely intact.

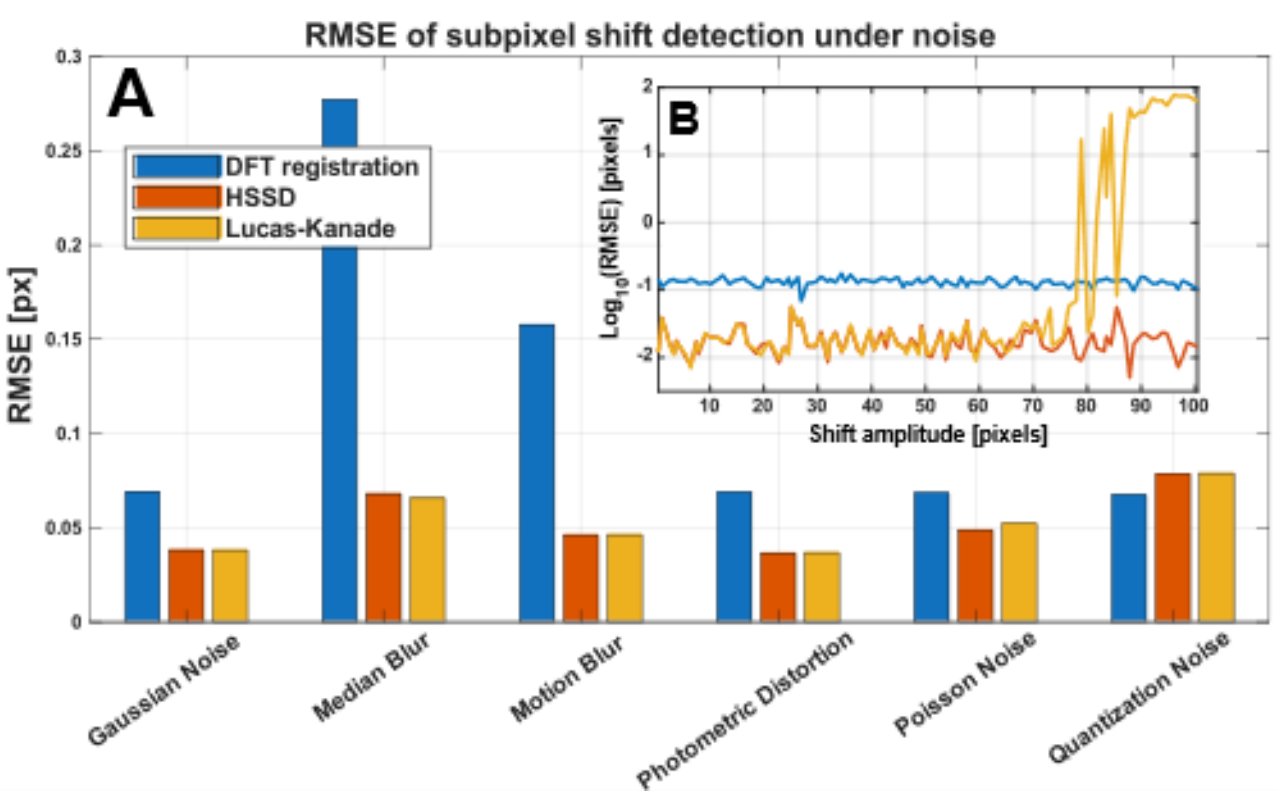

Fig. 2. RMSE comparison of subpixel shift detection methods (DFT registration, HSSD, and LK) under different noise types for narrow displacement range. Panel A presents absolute RMSE values per noise category; panel B shows $\log_{10}$(RMSE) as a function of displacement amplitude under affine shift. The results highlight the superior robustness of LK and HSSD against photometric and stochastic noise, particularly in scenarios involving blur distortions where DFT precision significantly declines. Conversely, DFT registration demonstrates

enhanced performance under quantization noise, slightly outperforming the differential methods in this specific category.

**Experimental validation and comparative analysis**. The experimental validation was performed using a DIHM setup as illustrated in Fig. 3. The system utilizes a coherent 532 nm laser illumination, enabling high-contrast diffraction patterns for both lateral and axial particle localization. The optical configuration features a high-numerical-aperture objective (100x, 0.95 NA), providing a total system magnification of 100 onto a Teledyne FLIR BFS-U3-120S4M-CS monochrome camera. With a physical pixel size of 1.85 µm and a system magnification of 100×, the pixel-to-physical-unit calibration coefficient is 18.5 nm/px, applied throughout to convert all measured pixel displacements to physical units. To facilitate precise motion control, the sample was mounted on a three-axis piezo-actuated stage (Piezoconcept LT3.100), which allows for independent and simultaneous positioning along the X, Y, and Z axes with a motion resolution of 0.1 nm and a full-range repeatability of 0.2 nm. The geometric relationship between the stage and camera coordinate systems was further characterized through a dedicated calibration procedure performed prior to each measurement session, in which the stage traversed a regular grid of known positions and the resulting image-plane displacements were recorded; an affine transformation matrix capturing pixel-to-distance scaling, rotational alignment between the axes, and residual shear was then fitted to the full dataset, confirming sub-pixel residual alignment errors across the full field of view. This nanometric precision, monitored by integrated high-resolution silicon sensors, enabled a rigorous investigation of sub-pixel detection stability under both planar translations and varying defocusing conditions, with the piezo stage serving as the displacement ground truth.

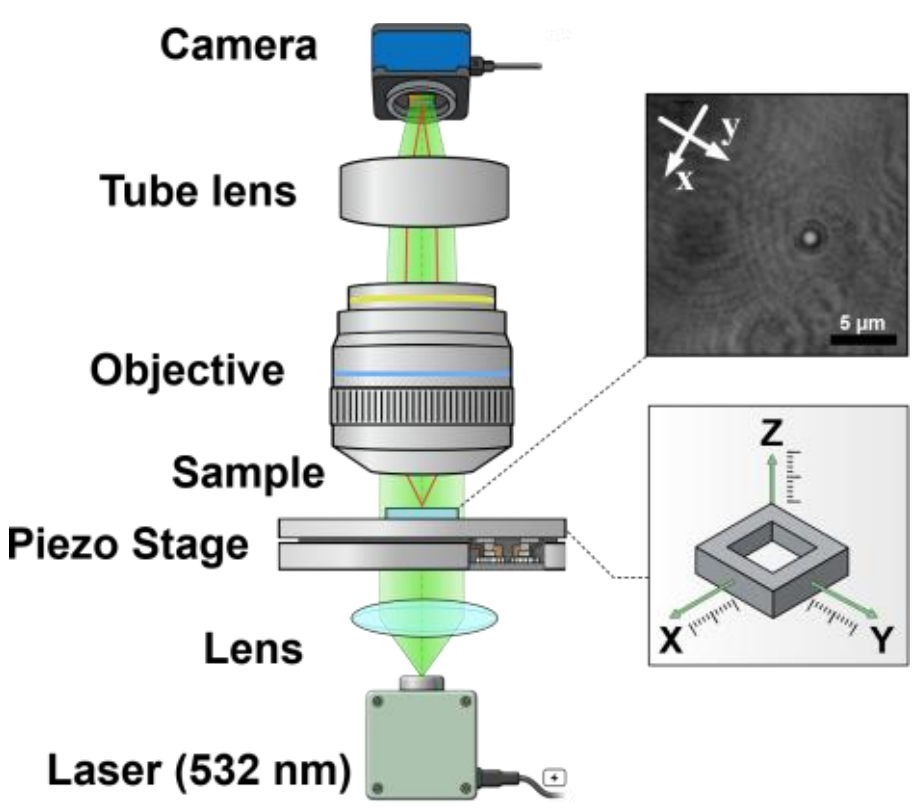

Fig. 3. Schematic diagram of the experimental setup, comprising a green laser point source, a sample mounted on a three-axis piezo stage, and a camera. The inset shows a representative in-line hologram of a microparticle recorded with the described system.

The performance of the DFT-based registration, the LK method, and the proposed HSSD approach was evaluated through a comparative analysis in the sub-pixel regime, Fig. 4. To establish a reliable ground truth, the piezoelectric stage was translated in 5 nm increments. We then compared these known physical displacements with the values estimated by each algorithm from the recorded images of a 1 µm (in XY) phase dot fabricated via two-photon polymerization. The structure was custom-fabricated using IP-Dip polymer resin (Nanoscribe), with a refractive index of 1.552 and a height of 0.4 µm, resulting in an approximate phase delay of 2.6 rad.

The results in Fig. 4 show that the standalone DFT method suffers from a "flattening" effect, failing to resolve the actual movement of the stage when the total shift remains below 5 pixels. While the LK method tracks these small increments more accurately, the HSSD approach provides the highest precision, closely following the physical displacement of the piezoelectric stage and outperforming both individual methods at this nanometer scale. The HSSD approach achieves the lowest RMSE of 1.21 pixels (22.4 nm), outperforming standalone DFT (1.44 px, 26.6 nm) and LK (1.89 px, 35.0 nm). These results confirm that the DFT initialization overcomes the limited capture range of LK, while the LK refinement eliminates the quantization errors of standalone frequency-domain methods.

It should be noted that the RMSE values reported above reflect contributions from physical noise sources inherent to the experimental setup - including mechanical vibrations and residual stage positioning uncertainty - rather than algorithmic limitations alone. As demonstrated by the numerical analysis, in which sub-pixel shifts were introduced computationally into real holographic images acquired with the same optical system, the intrinsic precision of all evaluated methods is genuinely sub-pixel, with HSSD achieving RMSE below 0.05 pixels across the majority of tested conditions. The experimental RMSE exceeding 1 pixel for all methods simultaneously indicates that system-level noise dominates at this scale, such that the experiment validates nanometric performance in physical units (HSSD: 22.4 nm) rather than serving as a sub-pixel ground truth in pixel units.

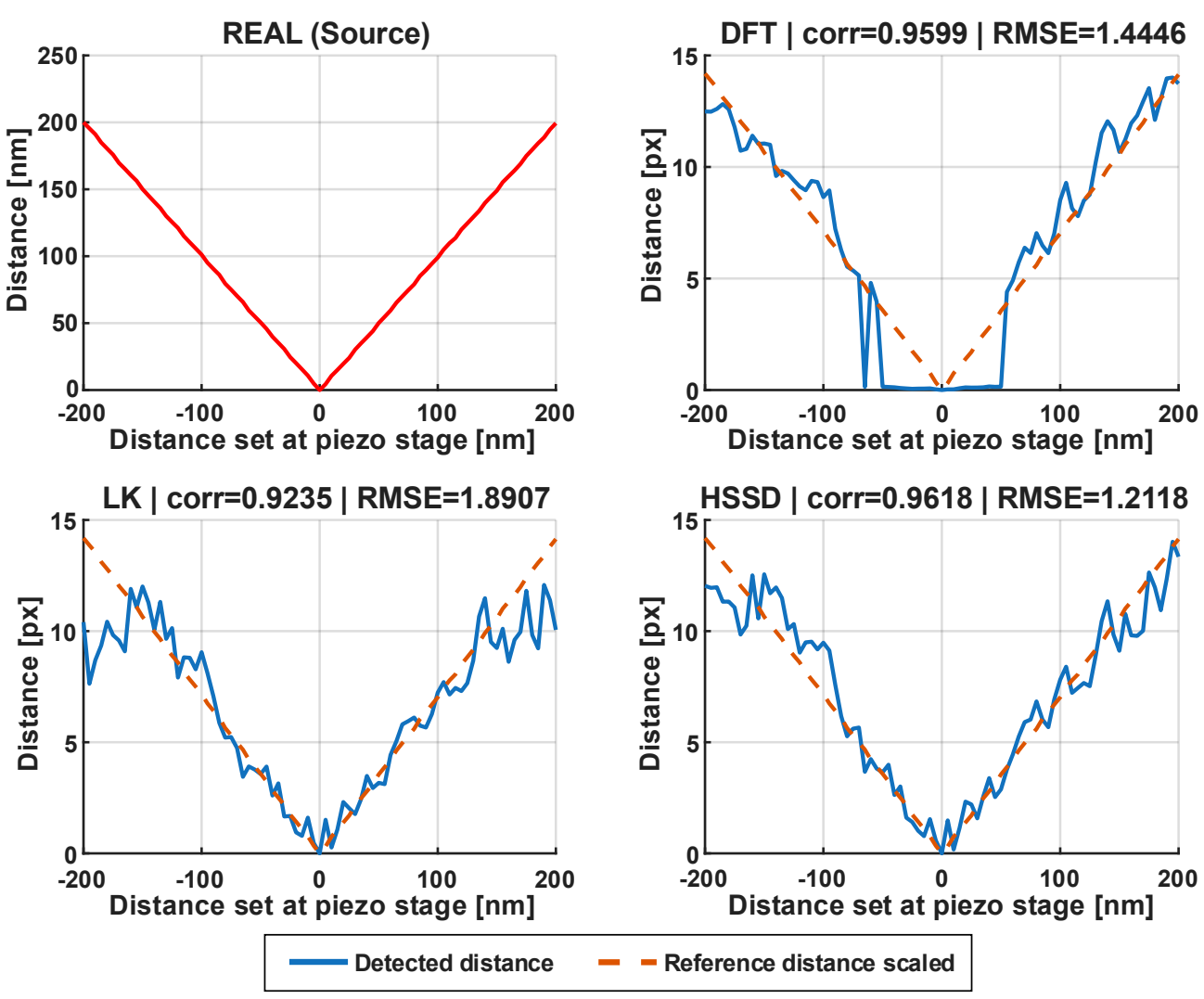

Fig. 4. Micro-displacement detection in 5 nm increments. Comparison of DFT, LK, and HSSD algorithms against the reference signal (REAL). Subplot titles include correlation coefficients and RMSE values for each method.

**Wide-range performance and XYZ tracking synergy.** DIHM enables full 3D object localization, as the recorded hologram encodes both lateral and axial information through the diffraction fringe morphology. While this work focuses on lateral XY tracking, the Z coordinate is estimated independently by DarkFocus [13], with neither estimator depending on the output of the other. The hybrid framework was evaluated using a sequence alternating between Y-axis translations of 0.2 µm and Z-axis displacements of 0.25 µm, Fig. 5. HSSD operates directly on the raw hologram without prior propagation, and the full XYZ trajectory is assembled by combining both estimates frame-by-frame. The near-zero XY signal observed during purely axial shifts confirms that HSSD correctly distinguishes defocus-induced fringe scaling from genuine lateral displacement, a robustness attributed to the DFT-based initialization anchoring the LK refinement to the dominant spatial frequency rather than local intensity.
The critical advantage of this strategy is evident when evaluating micron-scale displacements, where standalone algorithms fail; specifically, the standalone LK algorithm diverges completely once displacements exceed approximately 50 pixels, resulting in a total loss of correlation (corr = -0.08) and a massive error (RMSE = 47.81 px). In contrast, HSSD overcomes this limitation by using the DFT stage for robust initialization, ensuring that the LK refinement always operates within its convergence basin. Despite this marginal trade-off at extreme ranges, the hybrid approach remains essential for preventing catastrophic divergence while providing superior precision in the sub-pixel regime, effectively resolving the full 3D trajectory by decoupling lateral and axial motion components for reliable nanometric metrology in real-world DIHM applications.

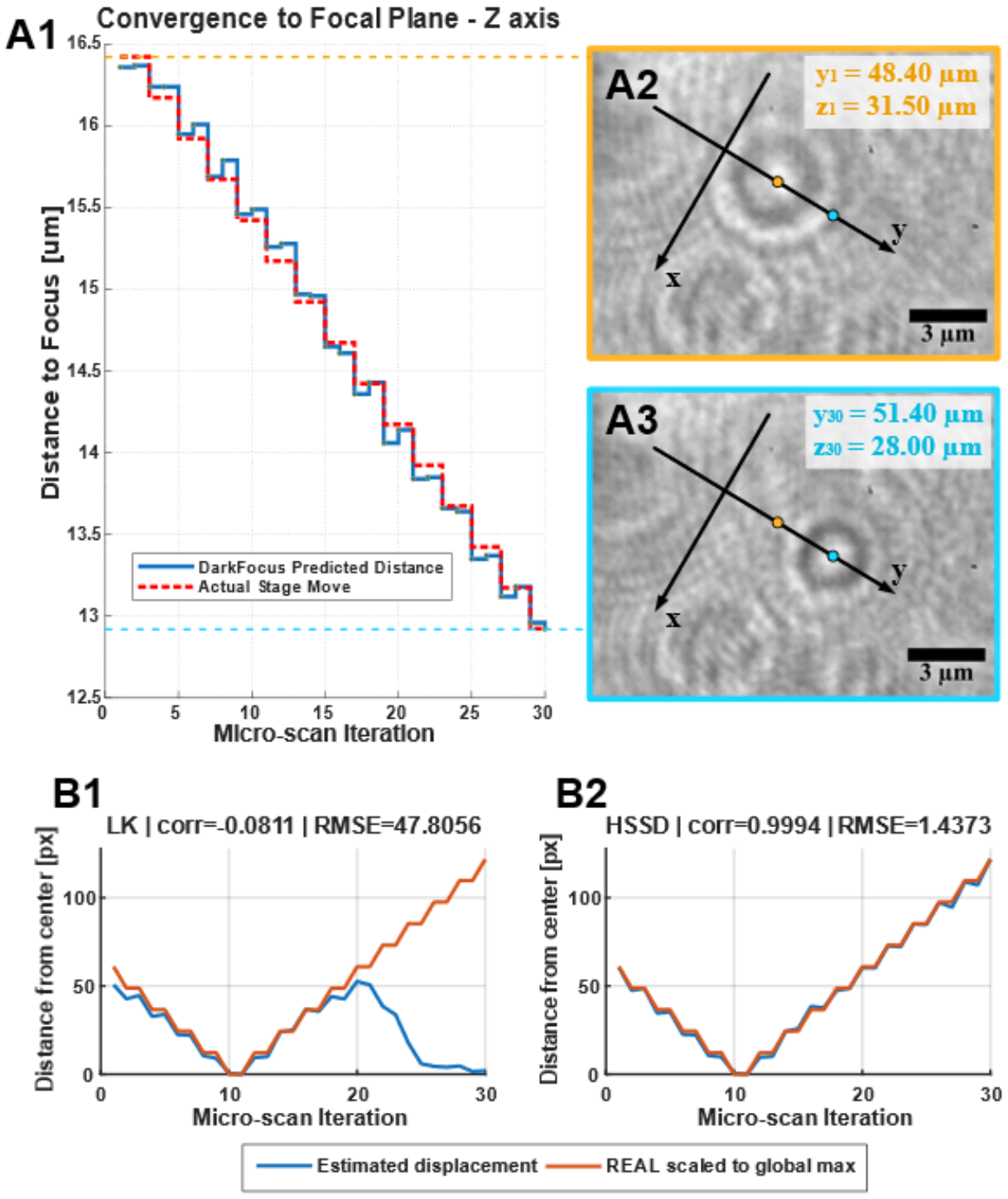

Fig. 5. Stepwise comparison of the DarkFocus predicted distance and the actual stage movement, aligned to a common reference frame. The scan follows a Y–Z stepping pattern with alternating lateral (Y) and axial (Z) increments. (A1) The axial (Z-axis) convergence plot, with small deviations indicating noise during nominally pure Y steps. (A2) and (A3) show holograms from the initial and final iteration, illustrating changes in particle position and diffraction fringe morphology. The lower plots present the lateral (Y-axis) displacement estimation: (B1) LK and (B2) HSSD compared against ground truth. HSSD remains consistent with the ground truth across the full range, while LK degrades for larger displacements, as reflected by reduced correlation and increased RMSE.

**In summary**, we have demonstrated the HSSD method - a DFT-LK framework - that leverages the complementary strengths of frequency- and spatial-domain motion estimation. This approach utilizes the frequency-domain stability of the DFT-based stage to provide a robust global initialization that effectively averages quantization and noise effects. This coarse estimate is critical to overcoming the inherent capture range limitations and initialization sensitivity of the LK algorithm, which fails at displacements exceeding approximately 50 pixels, as demonstrated in this work. Conversely, the LK stage provides the high-precision spatial refinement necessary to eliminate the sub-pixel resolution artifacts and interpolation errors typical of standalone DFT estimators. Both extensive numerical simulations and real-world experimental validation consistently confirm the superiority of this hybrid architecture: in simulation, HSSD achieves RMSE below 0.05 pixels across the majority of tested noise conditions; experimentally, HSSD achieves an RMSE of 1.21 px (22.4 nm at 18.5 nm/px effective sampling), outperforming standalone DFT (1.44 px) and LK (1.89 px), and successfully resolves displacement steps of 5 nm (≈0.27 px) where isolated methods fail. This synergy between global robustness and local gradient precision provides a reliable foundation for label-free particle tracking applications in digital holographic microscopy and beyond.

**Funding.** Funded by the European Union (ERC, NaNoLens, Project 101117392). Views and opinions expressed are however those of the author(s) only and do not necessarily reflect those of the European Union or the European Research Council Executive Agency (ERCEA). Neither the European Union nor the granting authority can be held responsible for them. Research was funded by the Warsaw University of Technology within the Excellence Initiative: Research University (IDUB) programme.